\pdfoutput=1

\documentclass[12pt,a4paper]{article}

\usepackage{ifthen} 
\newboolean{pdflatex}
\setboolean{pdflatex}{true} 

\newboolean{articletitles}
\setboolean{articletitles}{true} 

\newboolean{uprightparticles}
\setboolean{uprightparticles}{false} 

\def\paperauthors{LHCb collaboration} 
\def\paperasciititle{Template for writing LHCb papers} 
\def\papertitle{First observation of the radiative decay $\LbLzGam$} 
\def\paperkeywords{{High Energy Physics}, {LHCb}} 
\def\papercopyright{\the\year\ CERN for the benefit of the LHCb collaboration} 
\def\paperlicence{CC-BY-4.0 licence}
\def\paperlicenceurl{https://creativecommons.org/licenses/by/4.0/}

\usepackage[top=1in, bottom=1.25in, left=1in, right=1in]{geometry}

\usepackage{microtype}
\usepackage{lineno}  
\usepackage{xspace} 
\usepackage{caption} 

\usepackage{booktabs} 

\usepackage{graphicx}  
\usepackage{color}
\usepackage{colortbl}
\graphicspath{{./figs/}} 
\DeclareGraphicsExtensions{.pdf,.PDF,png,.PNG}

\usepackage{amsmath} 
\usepackage{amssymb}
\usepackage{amsfonts}
\usepackage{upgreek} 

\newcommand*\patchAmsMathEnvironmentForLineno[1]{%
\expandafter\let\csname old#1\expandafter\endcsname\csname #1\endcsname
\expandafter\let\csname oldend#1\expandafter\endcsname\csname
end#1\endcsname
 \renewenvironment{#1}%
   {\linenomath\csname old#1\endcsname}%
   {\csname oldend#1\endcsname\endlinenomath}%
}
\newcommand*\patchBothAmsMathEnvironmentsForLineno[1]{%
  \patchAmsMathEnvironmentForLineno{#1}%
  \patchAmsMathEnvironmentForLineno{#1*}%
}
\AtBeginDocument{%
\patchBothAmsMathEnvironmentsForLineno{equation}%
\patchBothAmsMathEnvironmentsForLineno{align}%
\patchBothAmsMathEnvironmentsForLineno{flalign}%
\patchBothAmsMathEnvironmentsForLineno{alignat}%
\patchBothAmsMathEnvironmentsForLineno{gather}%
\patchBothAmsMathEnvironmentsForLineno{multline}%
\patchBothAmsMathEnvironmentsForLineno{eqnarray}%
}

\usepackage{hyperxmp}

\usepackage[pdftex,
            pdfauthor={\paperauthors},
            pdftitle={\paperasciititle},
            pdfkeywords={\paperkeywords},
            pdfcopyright={Copyright (C) \papercopyright},
            pdflicenseurl={\paperlicenceurl}]{hyperref}

\usepackage[colorinlistoftodos,textsize=scriptsize]{todonotes}

\usepackage[all]{hypcap} 

\usepackage{xspace} 
\usepackage{upgreek}

\def\lhcb   {\mbox{LHCb}\xspace}

\def\babar  {\mbox{BaBar}\xspace}
\def\belle  {\mbox{Belle}\xspace}

\def\cdf    {\mbox{CDF}\xspace}

\def\MagUp {\mbox{\em Mag\kern -0.05em Up}\xspace}

\ifthenelse{\boolean{uprightparticles}}%
{
 
 \def\Pgamma      {\ensuremath{\upgamma}\xspace}

 \def\Peta        {\ensuremath{\upeta}\xspace}

 \def\Pmu         {\ensuremath{\upmu}\xspace}

 \def\Ppi         {\ensuremath{\uppi}\xspace}

 \def\Ppsi        {\ensuremath{\uppsi}\xspace}

 \def\PDelta      {\ensuremath{\Delta}\xspace}                 
 \def\PXi         {\ensuremath{\Xi}\xspace}                 
 \def\PLambda     {\ensuremath{\Lambda}\xspace}                 
 \def\PSigma      {\ensuremath{\Sigma}\xspace}                 
 \def\POmega      {\ensuremath{\Omega}\xspace}                 
 \def\PUpsilon    {\ensuremath{\Upsilon}\xspace}

 \def\PB      {\ensuremath{\mathrm{B}}\xspace}                 
                  
 \def\PD      {\ensuremath{\mathrm{D}}\xspace}

 \def\PJ      {\ensuremath{\mathrm{J}}\xspace}                 
 \def\PK      {\ensuremath{\mathrm{K}}\xspace}

 \def\Pb      {\ensuremath{\mathrm{b}}\xspace}                 
 \def\Pc      {\ensuremath{\mathrm{c}}\xspace}

 \def\Pi      {\ensuremath{\mathrm{i}}\xspace}

 \def\Pp      {\ensuremath{\mathrm{p}}\xspace}

 \def\Ps      {\ensuremath{\mathrm{s}}\xspace}

 \def\thebaroffset{0.0em}
}
{
 
 \def\Pgamma      {\ensuremath{\gamma}\xspace}

 \def\Peta        {\ensuremath{\eta}\xspace}

 \def\Pmu         {\ensuremath{\mu}\xspace}

 \def\Ppi         {\ensuremath{\pi}\xspace}

 \def\Ppsi        {\ensuremath{\psi}\xspace}                 
                  
 \mathchardef\PDelta="7101
 \mathchardef\PXi="7104
 \mathchardef\PLambda="7103
 \mathchardef\PSigma="7106
 \mathchardef\POmega="710A
 \mathchardef\PUpsilon="7107
                  
 \def\PB      {\ensuremath{B}\xspace}                 
                  
 \def\PD      {\ensuremath{D}\xspace}

 \def\PJ      {\ensuremath{J}\xspace}                 
 \def\PK      {\ensuremath{K}\xspace}

 \def\Pb      {\ensuremath{b}\xspace}                 
 \def\Pc      {\ensuremath{c}\xspace}

 \def\Pi      {\ensuremath{i}\xspace}

 \def\Pp      {\ensuremath{p}\xspace}

 \def\Ps      {\ensuremath{s}\xspace}

 \def\thebaroffset{0.18em}
}
\newcommand{\offsetoverline}[2][\thebaroffset]{\kern #1\overline{\kern -#1 #2}}%

\makeatletter
\ifcase \@ptsize \relax
  \newcommand{\miniscule}{\@setfontsize\miniscule{4}{5}}
\or
  \newcommand{\miniscule}{\@setfontsize\miniscule{5}{6}}
\or
  \newcommand{\miniscule}{\@setfontsize\miniscule{5}{6}}
\fi
\makeatother

\DeclareRobustCommand{\optbar}[1]{\shortstack{{\miniscule (\rule[.5ex]{1.25em}{.18mm})}
  \\ [-.7ex] $#1$}}

\def\mumu       {{\ensuremath{\Pmu^+\Pmu^-}}\xspace}

\def\g      {{\ensuremath{\Pgamma}}\xspace}

\def\squark    {{\ensuremath{\Ps}}\xspace}

\def\cquark    {{\ensuremath{\Pc}}\xspace}

\def\bquark    {{\ensuremath{\Pb}}\xspace}

\def\pion   {{\ensuremath{\Ppi}}\xspace}
\def\piz    {{\ensuremath{\pion^0}}\xspace}
\def\pip    {{\ensuremath{\pion^+}}\xspace}
\def\pim    {{\ensuremath{\pion^-}}\xspace}

\def\kaon    {{\ensuremath{\PK}}\xspace}

\def\KorKbar {\kern \thebaroffset\optbar{\kern -\thebaroffset \PK}{}\xspace}

\def\Kp      {{\ensuremath{\kaon^+}}\xspace}
\def\Km      {{\ensuremath{\kaon^-}}\xspace}

\def\KS      {{\ensuremath{\kaon^0_{\mathrm{S}}}}\xspace}

\def\Kstarz  {{\ensuremath{\kaon^{*0}}}\xspace}

\newcommand{\etaz}{\ensuremath{\Peta}\xspace}

\def\D       {{\ensuremath{\PD}}\xspace}

\def\DorDbar {\kern \thebaroffset\optbar{\kern -\thebaroffset \PD}\xspace}
\def\Dz      {{\ensuremath{\D^0}}\xspace}

\def\B       {{\ensuremath{\PB}}\xspace}

\def\BorBbar {\kern \thebaroffset\optbar{\kern -\thebaroffset \PB}\xspace}
\def\Bz      {{\ensuremath{\B^0}}\xspace}

\def\Bu      {{\ensuremath{\B^+}}\xspace}

\def\Bd      {{\ensuremath{\B^0}}\xspace}
\def\Bs      {{\ensuremath{\B^0_\squark}}\xspace}

\def\jpsi     {{\ensuremath{{\PJ\mskip -3mu/\mskip -2mu\Ppsi\mskip 2mu}}}\xspace}

\def\Y#1S{\ensuremath{\PUpsilon{(#1S)}}\xspace}

\def\proton      {{\ensuremath{\Pp}}\xspace}

\def\Lz          {{\ensuremath{\PLambda}}\xspace}

\def\LorLbar     {\kern \thebaroffset\optbar{\kern -\thebaroffset \PLambda}\xspace}

\def\Lc          {{\ensuremath{\Lz^+_\cquark}}\xspace}

\def\Lb           {{\ensuremath{\Lz^0_\bquark}}\xspace}

\def\BF         {{\ensuremath{\mathcal{B}}}\xspace}
\def\BR         {\BF}

\newcommand{\decay}[2]{\mbox{\ensuremath{#1\!\to #2}}\xspace}         

\def\to                 {\ensuremath{\rightarrow}\xspace}

\def\BsPhiGam     {\decay{\Bs}{\phi \g}}
\def\BdKstGam     {\decay{\Bd}{\Kstarz \g}}

\def\AT#1     {\ensuremath{A_{\mathrm{T}}^{#1}}\xspace}           
\def\btosgam  {\decay{\bquark}{\squark \g}}

\def\C#1      {\ensuremath{\mathcal{C}_{#1}}\xspace}                       
\def\Cp#1     {\ensuremath{\mathcal{C}_{#1}^{'}}\xspace}                    
\def\Ceff#1   {\ensuremath{\mathcal{C}_{#1}^{\mathrm{(eff)}}}\xspace}        
\def\Cpeff#1  {\ensuremath{\mathcal{C}_{#1}^{'\mathrm{(eff)}}}\xspace}       
\def\Ope#1    {\ensuremath{\mathcal{O}_{#1}}\xspace}                       
\def\Opep#1   {\ensuremath{\mathcal{O}_{#1}^{'}}\xspace}                    

\newcommand{\nospaceunit}[1]{\ensuremath{\text{#1}}}       
\newcommand{\aunit}[1]{\ensuremath{\text{\,#1}}}       

\newcommand{\tev}{\aunit{Te\kern -0.1em V}\xspace}
\newcommand{\gev}{\aunit{Ge\kern -0.1em V}\xspace}
\newcommand{\mev}{\aunit{Me\kern -0.1em V}\xspace}
\newcommand{\kev}{\aunit{ke\kern -0.1em V}\xspace}
\newcommand{\ev}{\aunit{e\kern -0.1em V}\xspace}
\newcommand{\mevc}{\ensuremath{\aunit{Me\kern -0.1em V\!/}c}\xspace}
\newcommand{\gevc}{\ensuremath{\aunit{Ge\kern -0.1em V\!/}c}\xspace}
\newcommand{\mevcc}{\ensuremath{\aunit{Me\kern -0.1em V\!/}c^2}\xspace}
\newcommand{\gevcc}{\ensuremath{\aunit{Ge\kern -0.1em V\!/}c^2}\xspace}

\def\mm   {\aunit{mm}\xspace}

\def\mum  {\ensuremath{\,\upmu\nospaceunit{m}}\xspace}

\def\fb   {\ensuremath{\aunit{fb}}\xspace}
\def\invfb   {\ensuremath{\fb^{-1}}\xspace}

\def\gsim{{~\raise.15em\hbox{$>$}\kern-.85em
          \lower.35em\hbox{$\sim$}~}\xspace}
\def\lsim{{~\raise.15em\hbox{$<$}\kern-.85em
          \lower.35em\hbox{$\sim$}~}\xspace}

\def\pt         {\ensuremath{p_{\mathrm{T}}}\xspace}

\def\ptot       {\ensuremath{p}\xspace}
\def\et         {\ensuremath{E_{\mathrm{T}}}\xspace}

\def\rad{\aunit{rad}}

\def\evtgen     {\mbox{\textsc{EvtGen}}\xspace}

\def\geant      {\mbox{\textsc{Geant4}}\xspace}

\def\photos     {\mbox{\textsc{Photos}}\xspace}

\def\pythia     {\mbox{\textsc{Pythia}}\xspace}

\xspace

\def\tell1  {TELL1\xspace}
\def\ukl1   {UKL1\xspace}

\newcommand{\eg}{\mbox{\itshape e.g.}\xspace}
\newcommand{\ie}{\mbox{\itshape i.e.}\xspace}

\def\splot {\textit{sPlot}\xspace}

\def\btosgam  {\decay{\bquark}{\squark\g}}

\def\LbLzGam  {\decay{\Lb}{\Lz \g}}

\def\LbpKGam  {\decay{\Lb}{\proton\Km \g}}

\def\BdKpipiz {\decay{\Bd}{\Kp \pim \piz}}

\def\BdKstzEta {\decay {\Bd} {\Kstarz \etaz}}

\def\BKpipiGam {\decay{\B}{\Kp\pim\pion\g}}
\def\BKpipizX {\decay{\B}{\Kp \pim \piz X}}

\def\LbtoLz {\decay{\Lb}{\Lz}}
\def\Lzppi {\decay{\Lz}{\proton\pim}}
\def\KstKpi {\decay{\Kstarz}{\Kp\pim}}
\def\LbLzJpsi {\decay{\Lb}{\jpsi\Lz}}

\def\LbpKJpsi {\decay{\Lb}{\jpsi\proton\Km}}
\def\LbLzEta {\decay{\Lb}{\Lz \etaz}}
\def\Etagg	{\decay{\etaz}{\g\g}}

\def\LbLzpiz {\decay{\Lb}{\Lz \piz}}

\def \DzKPi          {\decay{\Dz}{\Km\pip}}
\def \LbLzmm      {\decay {\Lb}{\Lz\mumu}}
\def\Kspipi {\decay{\KS}{\pip\pim}}

\def\BRLbLzGam {\ensuremath{\BF(\LbLzGam)}\xspace}
\def\BRBdKstGam {\ensuremath{\BF(\BdKstGam)}\xspace}

\def\BRLbLzmm {\ensuremath{\BF(\LbLzmm)}\xspace}

\def\BRLzppi {\ensuremath{\BF(\Lzppi)}\xspace}
\def\BRKstKpi {\ensuremath{\BF(\KstKpi)}\xspace}

\usepackage{cite} 
\usepackage{mciteplus}

\usepackage{longtable} 

\begin{document}

\renewcommand{\thefootnote}{\fnsymbol{footnote}}
\setcounter{footnote}{1}


\begin{titlepage}
\pagenumbering{roman}

\vspace*{-1.5cm}
\centerline{\large EUROPEAN ORGANIZATION FOR NUCLEAR RESEARCH (CERN)}
\vspace*{1.5cm}
\noindent
\begin{tabular*}{\linewidth}{lc@{\extracolsep{\fill}}r@{\extracolsep{0pt}}}
\ifthenelse{\boolean{pdflatex}}
{\vspace*{-1.5cm}\mbox{\!\!\!\includegraphics[width=.14\textwidth]{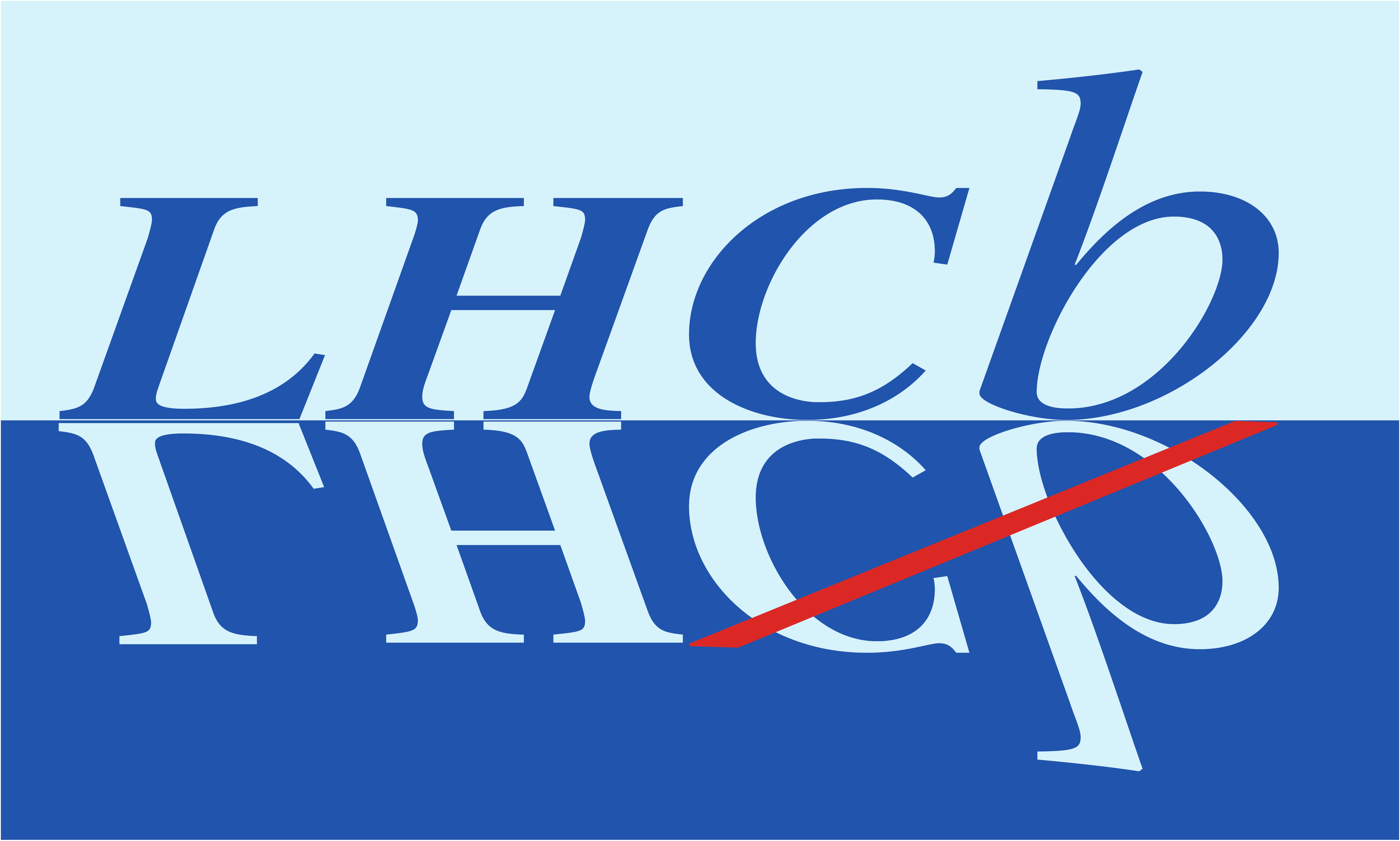}} & &}%
{\vspace*{-1.2cm}\mbox{\!\!\!\includegraphics[width=.12\textwidth]{figs/lhcb-logo.eps}} & &}%
\\
 & & CERN-EP-2019-060 \\  
 & & LHCb-PAPER-2019-010 \\  
 & & July 24, 2019 \\ 
 & & \\
\end{tabular*}

\vspace*{4.0cm}

{\normalfont\bfseries\boldmath\huge
\begin{center}
  \papertitle 
\end{center}
}

\vspace*{2.0cm}

\begin{center}
\paperauthors\footnote{Authors are listed at the end of this Letter.}
\end{center}

\vspace{\fill}

\begin{abstract}
  \noindent
  The radiative decay \decay{\Lb}{\Lz\gamma} is observed for the first time using a data sample of proton-proton collisions corresponding to an integrated luminosity of $1.7\,\invfb$
  collected by the \lhcb experiment at a center-of-mass energy of $13\,\tev$.
  Its branching fraction is measured exploiting the \BdKstGam decay as a normalization mode and is found to be
  \mbox{$\BRLbLzGam = (7.1 \pm 1.5 \pm 0.6 \pm 0.7) \times 10^{-6}$}, where the quoted uncertainties are statistical, systematic and systematic from external inputs, respectively.
  This is the first observation of a radiative decay of a beauty baryon.
\end{abstract}

\vspace*{2.0cm}

\begin{center}
  Published in Phys. Rev. Lett. 123 (2019) 031801
\end{center}

\vspace{\fill}

{\footnotesize 
\centerline{\copyright~\papercopyright. \href{\paperlicenceurl}{\paperlicence}.}}
\vspace*{2mm}

\end{titlepage}


\newpage
\setcounter{page}{2}
\mbox{~}

\cleardoublepage


\renewcommand{\thefootnote}{\arabic{footnote}}
\setcounter{footnote}{0}



\pagestyle{plain} 
\setcounter{page}{1}
\pagenumbering{arabic}



The decay $\Lb \to \Lz \gamma$ proceeds via the $b\to s\gamma$ flavour-changing neutral-current transition. This process is forbidden at tree level in the Standard Model (SM) and is therefore sensitive to new particles entering the loop-level transition, which can modify decay properties.
The polarization of the photon in these processes is predicted to be predominantly left-handed in the SM, up to small corrections of the order $m_{\squark}/m_{\bquark}$~\cite{Atwood:1997zr}.
While precise measurements of branching fractions and charge-parity-violation observables in radiative \bquark-meson decays previously performed at the \babar, \belle and \lhcb collaborations~\cite{Lees:2012ym, Saito:2014das, Horiguchi:2017ntw, LHCb-PAPER-2012-019}
are in agreement with SM calculations~\cite{Bosch:2001gv, Beneke:2004dp, Grinstein:2000pc, Keum:2004is, Ball:2006eu, Misiak:2015xwa, Paul:2016urs},
they do not provide stringent constraints on the presence of right-handed contributions to \btosgam transitions~\cite{Ushiroda:2006fi, Aubert:2008gy, LHCb-PAPER-2014-066, LHCb-PAPER-2016-034}.
Radiative \bquark-baryon decays have never been observed and offer a unique benchmark to measure the photon polarization due to the non-zero spin of the initial- and final-state particles~\cite{Gremm:1995nx}.
In particular, the \LbLzGam decay has been proposed as a suitable mode for the study of the photon polarization,
since the helicity of the \Lz baryon can be measured, giving access to the helicity structure of the \btosgam transition~\cite{Mannel:1997xy,Hiller:2001zj}.

The \LbLzGam decay is experimentally challenging to reconstruct. At high-energy hadron colliders the \Lb decay vertex cannot be determined directly due to the
long lifetime of the weakly decaying \Lz baryon and the unknown photon direction, when reconstructed as a cluster in the electromagnetic calorimeter.
Photons converting to a pair of electrons in the detector material could be used to reconstruct the photon direction but at the cost of a large efficiency loss.
This approach was used by the \cdf experiment to set the best limit on the branching fraction of this decay, \mbox{$\BRLbLzGam < 1.3 \times 10^{-3}$} at $90\%$ CL~\cite{Acosta:2002fh}.
This measurement still leaves ample room for improvement before achieving a sensitivity comparable to the SM prediction of \BRLbLzGam, which lies in the range
\mbox{(6--100)$\times 10^{-7}$},
where the large variation is due to different computations of the \LbtoLz form factors at the photon pole~\cite{Wang:2008sm, Mannel:2011xg, Gan:2012tt, Gutsche:2013pp, Faustov:2017wbh}.
A precise measurement of the branching fraction of this decay allows discrimination between different approaches to the form-factor computation, and is an important step towards the measurement of the photon polarization in radiative \bquark-baryon decays.

The \lhcb experiment provides unique conditions to study the \LbLzGam mode thanks to the large production of \Lb baryons at the LHC~\cite{LHCb-PAPER-2016-031,LHCb-PAPER-2018-050} 
and the excellent properties of the detector optimized for the analysis of \bquark-hadron decays. This Letter presents the first observation of the 
\LbLzGam decay, with \Lz reconstructed as \Lzppi, by the \lhcb experiment. The well-known radiative decay \BdKstGam~\cite{HFLAV16} is used as a normalization mode to measure the \LbLzGam branching fraction.
The data sample used in this work corresponds to $1.7\,\invfb$ of integrated luminosity collected by the \lhcb experiment in $13\tev$ proton-proton ($pp$) collisions during 2016. 
The results were not inspected until all analysis procedures were finalised.


The \lhcb detector~\cite{Alves:2008zz,LHCb-DP-2014-002} is a single-arm forward spectrometer covering the \mbox{pseudorapidity} range $2<\eta <5$.
The detector includes a high-precision tracking system consisting of a silicon-strip vertex detector surrounding the $pp$ interaction region, a large-area silicon-strip detector located upstream of a dipole magnet with a bending power of about $4{\mathrm{\,Tm}}$, and three stations of silicon-strip detectors and straw drift tubes placed downstream of the magnet.
The tracking system provides a measurement of the momentum, \ptot, of charged particles with a relative uncertainty that varies from 0.5\% at low momentum to 1.0\% at 200\gev.\footnote{Natural units with $\hbar=c=1$ are used throughout, so that mass and momentum are measured in units of energy.}
The minimum distance of a track to a primary vertex (PV), is measured with a resolution of $(15+29/\pt)\mum$, where \pt is the component of the momentum transverse to the beam, in\,\gev.
Different types of charged hadrons are distinguished using information from two ring-imaging Cherenkov detectors. 
Photons, electrons and hadrons are identified by a calorimeter system consisting of scintillating-pad and preshower detectors, an electromagnetic and a hadronic calorimeter.
Charged and neutral clusters in the electromagnetic calorimeter are separated by extrapolating the tracks reconstructed by the tracking system to the calorimeter plane, while photons and neutral pions
are distinguished by cluster shape and energy distributions.
For decays with high-energy photons in the final state, such as \BdKstGam,
a \Bz mass resolution around $100\,\mev$ is achieved~\cite{LHCb-PAPER-2011-042,LHCb-PAPER-2016-034},
dominated by the photon energy resolution.
The online event selection is performed by a trigger, which consists of a hardware stage, based on information from the calorimeter and muon systems, followed by a software stage, which applies a full event reconstruction.

At the hardware-trigger stage, events are required to have a cluster in the electromagnetic calorimeter with transverse energy \et above a threshold that varies in the range $2.1-3.0\,\gev$.
The software trigger requires at least one charged particle to have transverse momentum $\pt > 1\gev$ and to be inconsistent with originating from any PV.
Finally, a vertex is formed with two tracks significantly displaced from any PV and the combination with a high-\et photon is used to identify
decays consistent with the signal and normalization modes.
In the offline selection, trigger signals are associated with reconstructed particles.
Only events in which the trigger was fired due to the signal candidate are kept.


Simulated events are used to model the effects of the 
detector acceptance and the imposed selection requirements. 
In the simulation, $pp$ collisions are generated using \pythia~\cite{Sjostrand:2007gs,*Sjostrand:2006za} with a specific \lhcb configuration~\cite{LHCb-PROC-2010-056}.
Decays of unstable particles are described by \evtgen~\cite{Lange:2001uf}, in which final-state radiation is generated using \photos~\cite{Golonka:2005pn}.
The interaction of the generated particles with the detector, and its response, are implemented using the \geant toolkit~\cite{Allison:2006ve, *Agostinelli:2002hh} as described in Ref.~\cite{LHCb-PROC-2011-006}.
The signal sample is generated with unpolarized \Lb and only a left-handed photon contribution.
The agreement between data and simulation is validated using the \LbpKJpsi, \LbLzJpsi and \BdKstGam control modes exploiting the selections described in 
Refs.~\cite{LHCb-PAPER-2016-049},~\cite{LHCb-PAPER-2012-057} and~\cite{LHCb-PAPER-2016-034}, respectively. The \Lb momentum distribution of all simulated samples involving \Lb decays is corrected for discrepancies between the data and simulation in two-dimensional bins of \Lb momentum and \pt, $\ptot(\Lb)$ and $\pt(\Lb)$, using \LbpKJpsi background-subtracted data and simulated candidates. 

Signal candidates are reconstructed from the combination of a \Lz baryon and a high-energy photon candidate.
Good-quality tracks, consistent with the proton and pion hypotheses, with opposite charge and well separated from any PV, are combined to form the \Lz candidate.
Proton and pion candidates are required have \pt larger than $800\mev$ and $300\mev$, respectively. 
The proton-pion system is required to have an invariant mass in the range $1110$--$1122\,\mev$ and to form a good vertex that is well separated from the nearest PV.
Only \Lz candidates that decay in the highly segmented part of the vertex detector ($z < 270 \mm$) and have a \pt larger than $1\gev$ are retained for further study. 
Photons, reconstructed from clusters in the electromagnetic calorimeter, must be consistent with originating from a neutral particle and have $\et > 3\gev$.
The photon direction is computed assuming it is produced in the interaction region. 
The sum of the \Lz \pt and the photon \et should be larger than $5\gev$.
The \Lb four-momentum is obtained as the sum of the \Lz and photon candidate four-momenta.
The \Lb transverse momentum is required to be above $4\,\gev$ and its invariant mass within $900 \mev$ of the known \Lb mass~\cite{PDG2018}.
Since the origin vertex of the photon is not known, the \Lb decay vertex is not reconstructed
and therefore
it is not possible to use its displacement with respect to the PV to separate background coming directly from the $pp$ collision.
Instead, the distance of closest approach (DOCA) between the \Lb and \Lz trajectories is required to be small,
where the former is calculated using the reconstructed momentum and assuming it originates at the PV closest to the \Lz trajectory.
Candidates for the normalization channel \BdKstGam are reconstructed following similar criteria.
In this case, tracks are required to be consistent with the \kaon and \pion hypotheses, their invariant mass must be within $100 \mev$ of the known \Kstarz mass~\cite{PDG2018}, and the \Bd candidate mass is required to be in the range $4600-6180 \mev$.

A Boosted Decision Tree (BDT)~\cite{Breiman}, employing the XGBoost algorithm~\cite{Chen:2016:XST:2939672.2939785} and implemented through 
the Scikit-learn library~\cite{Scikit-learn-paper}, is used to further separate signal from combinatorial background. It is trained on simulated events as proxy to the signal and on 
data candidates with an invariant mass larger than $6.1\gev$ as background. A combination of topological and isolation information is used as input for the classifier,
including the transverse momentum and the separation from the PV of the different particles, the separation between the \Lz decay vertex and the PV and the DOCA between the two tracks and between the \Lb and \Lz trajectories.
Background \Lb candidates with extra tracks close to the \Lz or photon candidates are rejected using the asymmetry of the sum of momenta of all the tracks present in a cone of $1\rad$ around the particle direction with respect to its momentum. Such tracks potentially arise from 
decays with additional particles in the final state that have not been reconstructed when building the \Lb candidate. 
A two-fold technique~\cite{kfold} is used to avoid overtraining and no correlation is observed between the BDT response and 
the candidate mass. The requirement on the BDT output is optimized using the Punzi figure of merit~\cite{Punzi:2003bu}. The chosen working point provides a background rejection of $99.8\%$ while retaining $33\%$ of the signal candidates.
A separate BDT with the same configuration and input variables is trained to select \BdKstGam candidates using 
simulated candidates as signal and data events in the high-mass sideband as background. In this case, the requirement on the BDT output is optimized by maximizing the signal significance
using the known branching fraction for this decay to compute the expected signal yield at each step.

Potential contamination from neutral pions that are reconstructed as a single merged cluster in the electromagnetic calorimeter is suppressed by employing a neural network classifier
trained to separate \piz mesons from photons.
This classifier exploits the broader shape of the calorimeter cluster of a \piz meson with respect to that of a single photon by using as input a
set of variables based on the combination of shower shape and energy information from the different calorimeter subsystems~\cite{CalvoGomez:2042173}.



The invariant-mass distribution of the selected candidates is used to disentangle signal from background through a maximum likelihood fit.
The \LbLzGam signal component is modeled with a double-tailed 
Crystal Ball~\cite{Skwarnicki:1986xj} probability density function (PDF), with power-law tails above and below the \Lb mass. 
The tail parameters are fixed to values determined from simulation while the mean and width of the signal peak are related to those of the \Bd meson
using simulation and the mass difference between the \Lb and \Bd hadrons measured by \lhcb~\cite{LHCb-PAPER-2014-002}. 
Several sources of background are investigated but only two are found to be significant.
The narrow width of the \Lz baryon~\cite{PDG2018} and the clean signature of the high-\pt proton
allow a pure hadronic selection, reducing the contamination from charged particle misidentification, \eg, coming from \Kspipi decays misidentified as \Lzppi candidates, to a negligible level.
Potentially dangerous backgrounds from decays with a similar topology to the signal and an additional pion have been studied and found to be negligible. Decays with intermediate \Lc states, like \decay{\Lb}{\Lc\pim} with \decay{\Lc}{\Lz\pip\piz}, 
are found to populate an invariant-mass range outside our fit region, and the topologically similar decay \LbLzpiz is expected to be suppressed 
due to the absence of QCD penguin contributions in this decay mode~\cite{Zhu:2018jet}.
The dominant source of background is formed by combinations of a real \Lz baryon with a random photon, referred to as combinatorial background,
and is modeled with an exponential PDF with a free decay parameter.
A small contamination from \LbLzEta decays with \Etagg, where one of the photons is not reconstructed, is also expected and is described with the shape determined from simulation.
The signal and combinatorial yields are free to float in the fit to data,
while the yield of \LbLzEta is constrained using the known branching fraction~\cite{PDG2018} and the reconstruction and selection efficiencies determined from simulation.

The mass distribution of \BdKstGam signal candidates is also described by a Crystal Ball function with two power-law tails with 
the parameters obtained from simulated events. The combinatorial component is modeled as an exponential PDF. 
Partially reconstructed backgrounds, \ie, background decays where one or more particles have not been reconstructed,
are copious in this case, mostly originating from the charged meson \Bu.
Three contributions are accounted for and modeled with shapes obtained from simulation:
two inclusive ones encompassing decays where one pion has not been reconstructed, referred to as \BKpipiGam, and decays with a neutral pion in the final state and any 
missing particle, referred to as \BKpipizX; and \BdKstzEta decays, where one of the photons from the \Etagg decay has not been reconstructed.
Backgrounds due to particle misidentification are also more abundant in this case,
due to the broad width of the \Kstarz meson~\cite{PDG2018}.
Contributions from \BsPhiGam, \LbpKGam and \BdKpipiz decays are described with the shapes obtained
from simulation. The yields of the signal, combinatorial and inclusive partially reconstructed
background are allowed to float in the fit, while those of the \BdKstzEta, \BsPhiGam, \LbpKGam and \BdKpipiz 
decays are fixed to the values obtained from simulation and the measured branching fractions~\cite{PDG2018, RivesMolina:2230991}. 
The fit stability is validated by performing pseudoexperiments with various signal yield hypotheses before proceeding with the final fit to data.
It is also checked that the extraction of the signal branching fraction is unbiased for branching fraction hypotheses at 
least as large as $3 \times 10^{-6}$.

\begin{figure}
	\centering
	\includegraphics[width=0.45\linewidth]{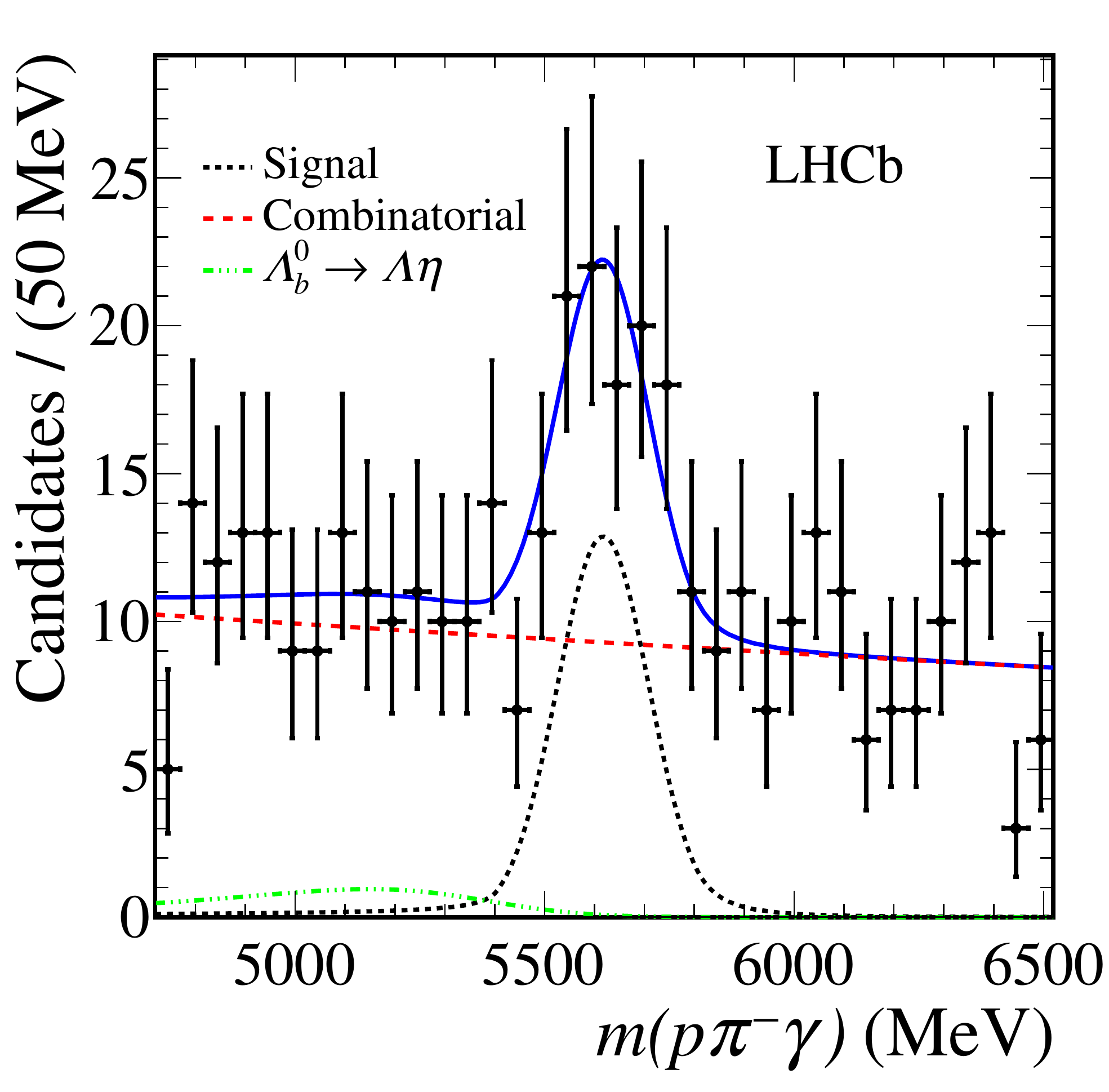}
	\includegraphics[width=0.45\linewidth]{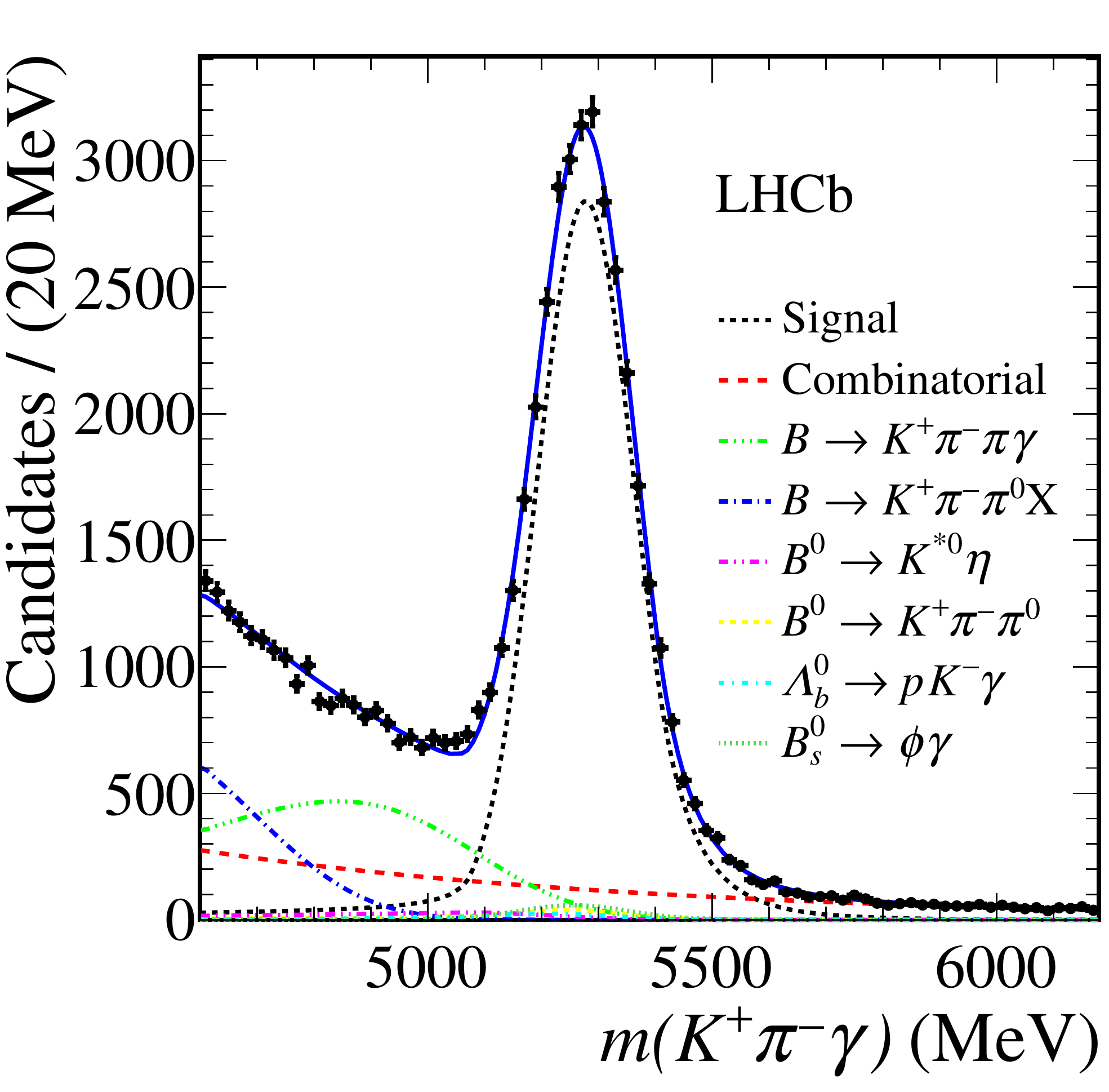}
	\caption{Simultaneous fit to the (left) \LbLzGam and (right) \BdKstGam invariant-mass 
		distributions of selected candidates.
		The data are represented by black dots and the result of the fit by a solid blue curve while 
		individual contributions are represented in different line styles (see legend). 
		\label{fig:fit}}
\end{figure}

The yield of signal and normalization events is obtained from a simultaneous extended unbinned maximum likelihood fit to data. The ratio of yields is given by the expression
\begin{equation}
\frac{N(\LbLzGam)}{N(\BdKstGam)} = 
\frac{f_{\Lb}}{f_{\Bd}} \times \frac{\BRLbLzGam}{\BRBdKstGam} \times 
\frac{\BRLzppi}{\BRKstKpi} \times \frac{\epsilon(\LbLzGam)}{\epsilon(\BdKstGam)},
\end{equation}
where $f_{\Lb}/f_{\Bd}$ is the ratio of hadronization fractions, $\BR$ is the branching fraction and $\epsilon$ is the combined reconstruction and selection efficiency for the given decay. 
The latter is obtained from simulation, except for the efficiencies related to charged particle 
identification requirements, which are determined from calibration samples of \Lzppi and \DzKPi~\cite{LHCb-DP-2018-001}.
The results of the simultaneous fit to data candidates are shown in Fig.~\ref{fig:fit}.
The signal yields are found to be $65 \pm 13$ and $32670 \pm 290$ for \LbLzGam and \BdKstGam, respectively.
The ratio of hadronization and branching fractions is measured to be
\begin{displaymath}
\frac{f_{\Lb}}{f_{\Bd}} \times \frac{\BRLbLzGam}{\BRBdKstGam} \times 
\frac{\BRLzppi}{\BRKstKpi} = (9.9 \pm 2.0) \times 10^{-2},
\end{displaymath}
where the uncertainty is statistical only.
To determine the signal branching fraction, 
the ratio of hadronization fractions, $f_{\Lb}/f_{\Bd}$, is computed from the \lhcb measurement of this quantity as a function of the \pt of the \bquark baryon~\cite{LHCb-PAPER-2018-050} 
and from the distribution of $\pt(\Lb)$ in the signal simulation. An average over \pt of the ratio of hadronization 
fractions of $f_{\Lb}/f_{\Bd} = 0.60 \pm 0.05$ is obtained for this analysis, where the uncertainty is derived from Ref.~\cite{LHCb-PAPER-2018-050}.
Taking the known branching fractions of the normalization mode and intermediate decays from Ref.~\cite{PDG2018}, the signal branching fraction is measured to be
\begin{displaymath}
\BRLbLzGam = (7.1 \pm 1.5) \times 10^{-6},
\end{displaymath}
where the uncertainty is statistical only.

Using the \splot~\cite{Pivk:2004ty} technique,  the absence of potential remaining backgrounds entering in the signal
component is cross-checked. In particular, the invariant mass of the $p\pi$ system and the output of the neural network classifier separating \piz mesons from photons
for background-subtracted data candidates are found to be compatible with the expected signal distributions.

\begin{table}[tbp]
	\caption{Dominant systematic uncertainties on the measurement of \BRLbLzGam.
	The uncertainties arising from external measurements are given separately.
		\label{tab:syst}}
	
	\centering
	\begin{tabular}{l r}		
		\toprule
		Source		 						& Uncertainty (\%)	\\
		\midrule
		Data/simulation agreement		    & $ 7.7 $ 		\\
		\Lb fit model						& $ 3.0 $		\\
		\BdKstGam backgrounds 	            & $ 2.7 $       \\
		Size of simulated samples	        & $ 1.7 $ 		\\
		Efficiency ratio 				    & $ 1.4 $		\\
		\midrule
		Sum in quadrature					& $ 9.0 $		\\
		\midrule
		$f_{\Lb}/f_{\Bd}$				    & $ 8.7 $		\\
		Input branching fractions 	        & $ 3.0 $		\\
		\midrule
		Sum in quadrature					& $ 9.2 $		\\
		\bottomrule
	\end{tabular}	
\end{table}

The dominant systematic uncertainties are listed in Table~\ref{tab:syst}. The largest contribution arises 
from the limited knowledge of the ratio of hadronization fractions, $f_{\Lb}/f_{\Bd}$. Potential remaining differences 
between data and simulation are evaluated by changing the requirement on the BDT output, recomputing 
the efficiencies and repeating the mass fit. 
Further systematic uncertainties come from the limited precision of the input branching fractions, the signal and normalization 
fit models, the finite simulation samples used to compute the selection efficiencies 
and other uncertainties associated to the extraction of the ratio of efficiencies, including the uncertainties 
on the corrections applied to the simulation and systematic effects on 
the extraction of the particle identification and hardware trigger efficiencies.

The \LbLzGam signal significance is evaluated from a profile likelihood using Wilks' theorem~\cite{Wilks:1938dza}
and is confirmed with pseudoexperiments. 
Including both statistical and systematic uncertainties, the \LbLzGam decay is observed with a significance of $5.6\sigma$.


To summarize, a search for the \bquark-baryon flavour-changing neutral-current radiative decay \LbLzGam 
is performed with a data sample corresponding to an integrated luminosity of $1.7\invfb$ 
collected in $pp$ collisions at a center-of-mass energy of $13\tev$ with the \lhcb detector. 
A signal of $65 \pm 13$ decays is observed with a significance of $5.6 \sigma$.
This is the first observation of this mode and represents 
the first step towards the study of the photon polarization in radiative decays of \bquark-baryons with a larger dataset. Exploiting the well-known 
\BdKstGam mode as a normalization channel, the branching fraction of the \LbLzGam decay is measured for 
the first time, $\BRLbLzGam = (7.1 \pm 1.5 \pm 0.6 \pm 0.7) \times 10^{-6}$, where the first uncertainty is 
statistical, the second systematic and the third is the systematic from external measurements.
Our result is in good agreement with the predictions from Refs.~\cite{Wang:2008sm}, \cite{Mannel:2011xg}
and \cite{Gutsche:2013pp},
which make use of Light Cone Sum Rules, the Heavy Quark Limit and the covariant constituent quark model,
respectively.
A more recent calculation~\cite{Faustov:2017wbh}, which relies on the
Relativistic Quark Model and is able to predict accurately the integrated \BRLbLzmm measured by 
\lhcb~\cite{LHCb-PAPER-2015-009}, is compatible with the rate of \LbLzGam, although no uncertainties on this calculation are available. Other predictions~\cite{Gan:2012tt} are further away from our result, which can be used as input to future revisions.

\section*{Acknowledgements}
%
%
\noindent We express our gratitude to our colleagues in the CERN
accelerator departments for the excellent performance of the LHC. We
thank the technical and administrative staff at the LHCb
institutes.
We acknowledge support from CERN and from the national agencies:
CAPES, CNPq, FAPERJ and FINEP (Brazil); 
MOST and NSFC (China); 
CNRS/IN2P3 (France); 
BMBF, DFG and MPG (Germany); 
INFN (Italy); 
NWO (Netherlands); 
MNiSW and NCN (Poland); 
MEN/IFA (Romania); 
MSHE (Russia); 
MinECo (Spain); 
SNSF and SER (Switzerland); 
NASU (Ukraine); 
STFC (United Kingdom); 
NSF (USA).
We acknowledge the computing resources that are provided by CERN, IN2P3
(France), KIT and DESY (Germany), INFN (Italy), SURF (Netherlands),
PIC (Spain), GridPP (United Kingdom), RRCKI and Yandex
LLC (Russia), CSCS (Switzerland), IFIN-HH (Romania), CBPF (Brazil),
PL-GRID (Poland) and OSC (USA).
We are indebted to the communities behind the multiple open-source
software packages on which we depend.
Individual groups or members have received support from
AvH Foundation (Germany);
EPLANET, Marie Sk\l{}odowska-Curie Actions and ERC (European Union);
ANR, Labex P2IO and OCEVU, and R\'{e}gion Auvergne-Rh\^{o}ne-Alpes (France);
Key Research Program of Frontier Sciences of CAS, CAS PIFI, and the Thousand Talents Program (China);
RFBR, RSF and Yandex LLC (Russia);
GVA, XuntaGal and GENCAT (Spain);
the Royal Society
and the Leverhulme Trust (United Kingdom);
Laboratory Directed Research and Development program of LANL (USA).

\addcontentsline{toc}{section}{References}
\bibliographystyle{LHCb}
\bibliography{main,standard,LHCb-PAPER,LHCb-CONF,LHCb-DP,LHCb-TDR}

\newpage
\centerline
{\large\bf LHCb Collaboration}
\begin
{flushleft}
\small
R.~Aaij$^{29}$,
C.~Abell{\'a}n~Beteta$^{46}$,
B.~Adeva$^{43}$,
M.~Adinolfi$^{50}$,
C.A.~Aidala$^{77}$,
Z.~Ajaltouni$^{7}$,
S.~Akar$^{61}$,
P.~Albicocco$^{20}$,
J.~Albrecht$^{12}$,
F.~Alessio$^{44}$
% ...

{\footnotesize \it

% ...
$ ^{7}$Universit{\'e} Clermont Auvergne, CNRS/IN2P3, LPC, Clermont-Ferrand, France\\
% ...
$ ^{12}$Fakult{\"a}t Physik, Technische Universit{\"a}t Dortmund, Dortmund, Germany\\
% ...
$ ^{20}$INFN Laboratori Nazionali di Frascati, Frascati, Italy\\
% ...
$ ^{29}$Nikhef National Institute for Subatomic Physics, Amsterdam, Netherlands\\
% ...
$ ^{43}$Instituto Galego de F{\'\i}sica de Altas Enerx{\'\i}as (IGFAE), Universidade de Santiago de Compostela, Santiago de Compostela, Spain\\
$ ^{44}$European Organization for Nuclear Research (CERN), Geneva, Switzerland\\
% ...
$ ^{46}$Physik-Institut, Universit{\"a}t Z{\"u}rich, Z{\"u}rich, Switzerland\\
% ...
$ ^{50}$H.H. Wills Physics Laboratory, University of Bristol, Bristol, United Kingdom\\
% ...
$ ^{61}$University of Cincinnati, Cincinnati, OH, United States\\
% ...
$ ^{63}$Syracuse University, Syracuse, NY, United States\\
% ...
$ ^{77}$University of Michigan, Ann Arbor, United States, associated to $^{63}$\\
% ...
}
\end{flushleft}

\end{document}